\documentclass{ws-mplb}
\usepackage{amsfonts,amssymb,amsmath}
\usepackage{graphicx}
\usepackage[super,sort,compress]{cite} % sort => 1,2,3,5; nosort => 1,3,2,5; compress => 1-3,5; nocompress => 1,2,3,5

\begin{document}

\markboth{Bing Wang}{The transport phenomenon of  inertia Brownian particles in a periodic potential with non-Gaussian noise}

\catchline{}{}{}{}{}
%
%%%%%%%%%%%%%%%%%%%%%%%%%%%%%%%%%%%%%%%%%%%%%%%%%%%%%%%%%%%%%%%%%%%%

\title{The transport phenomenon of  inertia Brownian particles in a periodic potential with non-Gaussian noise}

\author{Bing Wang}
\address{Department of Mechanics and Physics, Anhui University of Science and Technology \\ Huainan, 232001, P.R.China\\
hnitwb@163.com}

\author{Xiaoxiao Zhang}

\address{Department of Mechanics and Physics, Anhui University of Science and Technology\\ Huainan, 232001, P.R.China}

\author{Yajuan Sun}

\address{Department of Mechanics and Physics, Anhui University of Science and Technology\\ Huainan, 232001, P.R.China}

\author{Zhongwei Qu}

\address{Department of Mechanics and Physics, Anhui University of Science and Technology\\ Huainan, 232001, P.R.China}

\author{Xuechao Li}
\address{Department of Mechanics and Physics, Anhui University of Science and Technology\\ Huainan, 232001, P.R.China}

\maketitle

\begin{history}
\received{(Day Month Year)}
\revised{(Day Month Year)}
\end{history}

\begin{abstract}
The transport phenomenon (movement and diffusion) of inertia Brownian particles in a periodic potential with non-Gaussian noise is investigated. It is found that proper noise intensity $Q$ will promote particles directional movement(or diffusion), but large $Q$ will inhibit this phenomenon. For large value of $Q$, the average velocity $\langle V\rangle$(or the diffusion coefficient $D$) has a maximum with increasing correlation time $\tau$. But for small value of $Q$, $\langle V\rangle$ (or $D$) decreases with increasing $\tau$.  In some cases, for the same value of $Q$ and the same value of $\tau$, non-Gaussian noise can induce particles directional movement(or diffusion), but Gaussian colored noise can not.
\end{abstract}

\section{Introduction}
Theory of Brownian motion has played a guiding role in the development of statistical physics. This theory provides a link between the microscopic dynamics and the observable macroscopic phenomena such as particles transport. Particles transport induced by zero average random perturbations plays a crucial role in many physical and biological systems, which has many theoretical and practical implications\cite{1,2,3,4,5,6}. Transport processes on microscale can exhibit entirely different properties from those encountered in the macroscopic world. The processes are strongly influenced by the presence of ubiquitous noise on microscale. This phenomenon has been predicted in different fields of physics, ranging from nano-devices to molecular motors\cite{7,8}. This working principle is a key for understanding processes of intracellular transport\cite{9}. Mahmud \emph{et al}. found random motions of motile cells can be rectified by asymmetric  microgeometries, interactions between the cells and imposed geometrical cues guide cell polarization and give rise to directional motility\cite{10}. Karnik \emph{et al}. demonstrated rectification of ionic transport in a nanofluidic diode fabricated by introducing a surface charge discontinuity in a nanofluidic channel\cite{11}.  Ai \emph{et al}. investigated the rectification and diffusion of nonintaracting self-propelled particles in a two dimensional corrugated channel and found the particles can be rectified by the self propelled velocity\cite{12}. Denisov \emph{et al}. investigated the transport properties of particles in spatially periodic structures which were driven by external time dependent forces manifestly depend on the spacetime symmetries of the corresponding equations of motion\cite{13}. Liao \emph{et al}. investigated the transport and diffusion of paramagnetic ellipsoidal particles under the action of a rotating magnetic field \cite{14}. As known, directed motion controllability has become a focal point of research in nonequilibrium statistical physics which inspired a plethora of new microscale devices displaying unusual transport features\cite{15,16,17,18,19,20,21,22,23,24,25,26,27,28}.

Most studies on noise induced particles transport phenomenon assumed that the noise source has a Gaussian distribution (either white or colored), and ignored the inertia term of the particle. However, some experimental results offer strong indications that the noise source could be non-Gaussian\cite{29,30}. Goswami \emph{et al}. studied the barrier crossing dynamics in the presence of the non-Gaussian noise and observed the multiplicative colored non-Gaussian noise could induce the resonant activation\cite{31}.

In the present paper, the transport phenomenon of inertia Brownian particles moving in a periodic potential with non-Gaussian noise is investigated. The paper is organized as follows: In Section \ref{label2}, the basic model of the system with a periodic potential and non-Gaussian noise is provided. In Section \ref{label3}, the effects of non-Gaussian noise is investigated by means of simulations.  In Section \ref{label4}, we get the conclusions.

\section{\label{label2}Basic model and methods}
In this work, we consider the generic ratchet model which consists of a classical inertial particle with mass $M$. The particle is governed by the following Langevin equation\cite{32}
\begin{equation}
M\frac{d^2x}{dt^2}+\gamma\frac{dx}{dt}=-U'(x)+A\cos (\Omega t)+\xi(t). \label{Mxt}
\end{equation}
$M\frac{d^2x}{dt^2}$ is the inertia term. $\gamma$ is the friction coefficient. $A\cos(\Omega t)$ is the unbiased time periodic force.  $A$ is the amplitude. $\Omega$ is the angular frequency. $U(x)$ is assumed to be in a double-sine form of period $2\pi L$
 \begin{equation}
U(x)=-\Delta U[\sin (\frac{x}{L})+\frac{1}{4}\sin (2\frac{x}{L}+\varphi -\frac{\pi}{2})].
\end{equation}
$\Delta U$ is the barrier height. The relative phase $\varphi$ between the two harmonics serves as a control parameter of the reflection-asymmetry of this potential. If $\varphi\neq0$ then generally its reflection symmetry is broken which we in turn classify as a ratchet-device.

$\xi(t)$ is a non-Gaussian noise with\cite{33}
\begin{equation}
\frac{d\xi(t)}{dt}=-\frac{1}{\tau}\frac{dV_q(\xi)}{d\xi}+\frac{1}{\tau}\varepsilon(t),
\end{equation}
$\varepsilon(t)$ is a Gaussian white noise of zero mean and correlation
$\langle\varepsilon(t)\varepsilon(t')\rangle=Q\delta(t-t')$. $V_q(\xi)$ is
\begin{equation}
V_q(\xi)=\frac{Q}{\tau(q-1)}\ln[1+\frac{\tau}{Q}(q-1)\frac{\xi^2}{2}], \label{VqQ}
\end{equation}
parameter $q$($|q-1|\leq1$) describes the deviation between $\xi(t)$ and Gaussian colored noise, and related to the Tsallis entropy\cite{33,34}. In the limit of $q\rightarrow 1$, process $\xi(t)$ coincides with the Gaussian colored noise with self-correlation time $\tau$ (Ornstein-Uhlenbeck process\cite{35}). If $q\neq1$, it is a non-Gaussian
noise term. $\tau$ is the correlation time of $\xi(t)$. $Q=\sqrt{2\gamma k_B T}$ is the noise intensity. $k_B$ is the Boltzmann constant. $T$ is the temperature of the heat bath.

Upon introducing characteristic length scale $L$, time scale $\frac{\gamma L^2}{\Delta U}$, and energy scale $\Delta U$, Eq. (\ref{Mxt}) can be rewritten in dimensionless form

\begin{equation}
m\frac{d^2\hat{x}}{d\hat{t}^2}+\frac{d\hat{x}}{d\hat{t}}=-\hat{U}'(\hat{x})+a\cos (\omega \hat{t})+\hat{\xi}(\hat{t}), \label{mxt}
\end{equation}
here, $\hat{x}=\frac{x}{L}$, $\hat{t}=\frac{t}{t_0}$, $t_0=\frac{\gamma L^2}{\Delta U}$. $\hat{U}(\hat{x})=U(x)/\Delta U=U(L\hat{x})/\Delta U=\hat{U}(\hat{x}+2\pi)$ possesses the period $2\pi$. Other parameters are: $m=M/(\gamma t_0)$, $a=AL/\Delta U$, $\omega=t_0 \Omega$. The rescaled thermal noise $\hat{\xi}(\hat{t})$ satisfies the following relation:
\begin{equation}
\frac{d\hat\xi(\hat{t})}{d\hat{t}}=-\frac{1}{\hat\tau}\frac{dV_q(\hat\xi)}{d\hat\xi}+\frac{1}{\hat\tau}\hat\varepsilon(\hat{t}),
\end{equation}
$\hat{\varepsilon}(\hat{t})$ is the Gaussian white noise with $\langle\hat{\varepsilon}(\hat{t})\rangle=0$ and $\langle\hat\varepsilon(\hat t)\hat\varepsilon(\hat t')\rangle=\hat{Q}\delta(\hat{t}-\hat{t'})$. The dimensionless noise intensity $\hat{Q} = k_BT/\Delta U$ is the ratio of thermal and the activation energy. In this paper, we will use the dimensionless equation (\ref{mxt}) and shall omit the hat ($\wedge$) for all quantities.

A central practical question in the theory of Brownian motors is the over all long-time behavior of the particles. The key quantities of particles transport in periodic potentials are the average velocity $\langle V\rangle$ and the diffusion coefficient $D$. The movement equation of Brownian particles in the present system can be described by the corresponding Fokker-Planck equation\cite{36,37}, but it is difficult to obtain the analytical expressions of $\langle V\rangle$ and $D$. $\langle V\rangle$ and $D$ can be corroborated by integration of the Langevin equations using the stochastic Euler algorithm. $\langle V\rangle$ can be obtained from the following equation:
\begin{equation}
\langle V\rangle=\lim_{t\to\infty}\frac{\langle{x(t)-x(t_0)}\rangle}{t-t_0},
\end{equation}
$x(t_0)$ is the position of particles at time $t_0$.

$D$ can be calculated by the formula
\begin{equation}
D=\lim_{t\rightarrow \infty}{\frac{\langle \Delta x^2(t)\rangle}{2t}}=\lim_{t\rightarrow \infty}{\frac{\langle x^2(t)\rangle-\langle x(t)\rangle^2}{2t}}.
\end{equation}
\section{\label{label3}Results and discussion}

In order to give a simple and clear analysis of the system. Eq.(\ref{mxt}) is integrated using the Euler algorithm with $\varphi=\frac{\pi}{2}$, $m=6$, $a=1.89$ and time step $\Delta t=10^{-3}$.

\begin{figure}
\center{
\includegraphics[height=8cm,width=10cm]{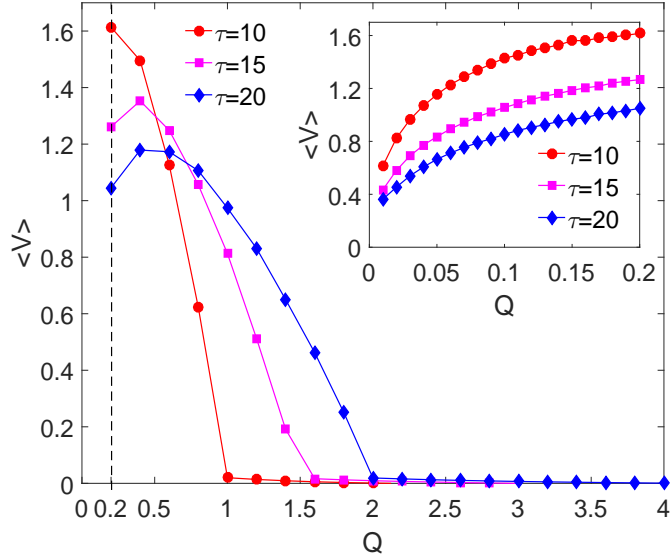}
\caption{Average velocity $\langle V\rangle$ as a function of noise intensity $Q$ for different values of correlation time $\tau$. The other parameters are $m=6$, $a=1.899$, $\omega=0.403$, $\varphi=\pi/2$, $q=0.8$.}\label{Fig1}
}
\end{figure}
Average velocity $\langle V\rangle$ as a function of the noise intensity $Q$($Q$ is the denominator in Eq.(\ref{VqQ}), so $Q\neq 0$) for different values of correlation time $\tau$ is shown in Fig.(\ref{Fig1}). It is found that average velocity $\langle V\rangle$ has a maximum with increasing $Q$. When $\tau=10$, $\langle V\rangle$ has a maximum $V_{max}\approx 1.614$ at $Q=0.2$. When $\tau=15$, $\langle V\rangle$ has a maximum $V_{max}\approx 1.353$ at $Q=0.4$. When $\tau=20$, $\langle V\rangle$ has a maximum $V_{max}\approx 1.179$ at $Q=0.4$. This means proper noise intensity $Q$ will promote particles movement. As $Q\rightarrow0$, the noise disappear, and the particles directional movement will disappear($\langle V\rangle\rightarrow 0$) too. The particles directional movement will disappear($\langle V\rangle\rightarrow 0$) if the noise intensity is too large($Q>2$).

\begin{figure}
\center{
\includegraphics[height=8cm,width=10cm]{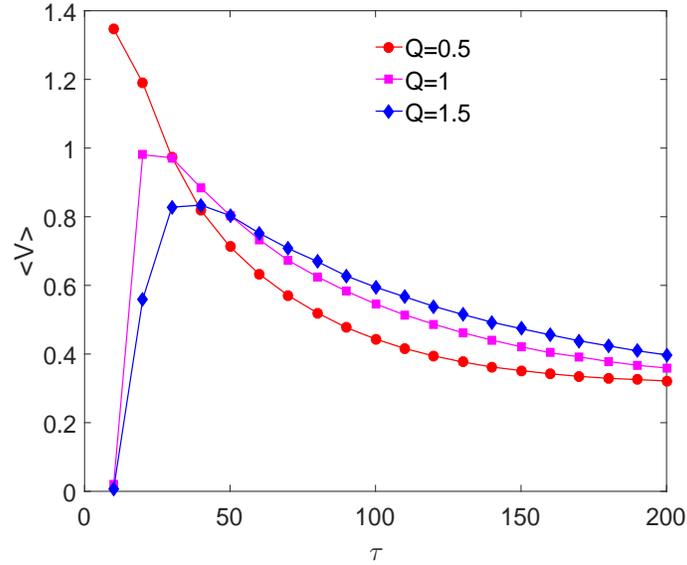}
\caption{Average velocity $\langle V\rangle$ as a function of correlation time $\tau$ for different values of noise intensity $Q$. The other parameters are $m=6$, $a=1.899$, $\omega=0.403$, $\varphi=\pi/2$, $q=0.8$.}\label{Fig2}
}
\end{figure}
Fig.\ref{Fig2} shows the average velocity $\langle V\rangle$  as a function of correlation time $\tau$($\tau$ is the denominator in Eq.(\ref{VqQ}), so $\tau\neq 0$) for different values of $Q$. When $Q=0.5$, $\langle V\rangle$ decreases monotonically with increasing $\tau$. When $Q=1$(or $Q=1.5$), $\langle V\rangle$ has a maximum with increasing $\tau$. So for small value of $Q$($Q=0.5$), the directional movement becomes indistinct with increasing $\tau$. But for large value of $Q$($Q=1$, $Q=1.5$), there exists an optimal value of $\tau$ at which the average velocity is maximal.

\begin{figure}
\center{
\includegraphics[height=8cm,width=10cm]{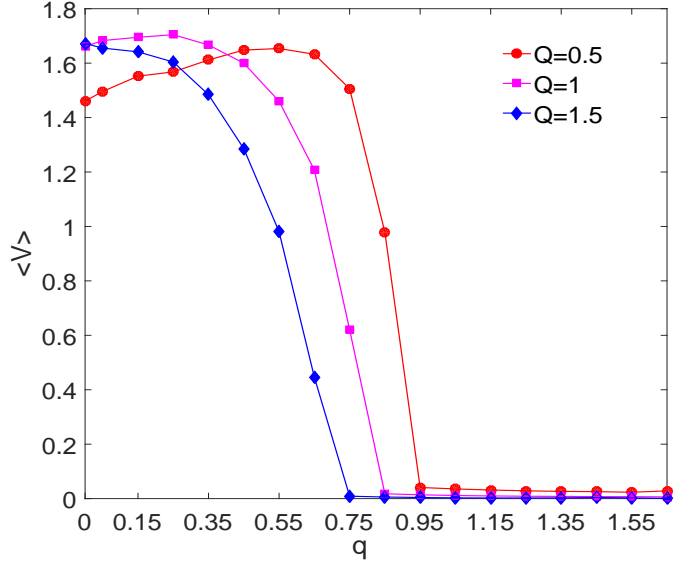}
\caption{Average velocity $\langle V\rangle$  as a function of $q$ for different values of $Q$. The other parameters are $\omega=0.403$, $\varphi=\pi/2$, $\tau=10$.}\label{Fig3}
}
\end{figure}
$\langle V\rangle$  as a function of the non-Guassian noise parameter $q$ with different values of noise intensity $Q$ is reported in Fig.(\ref{Fig3}). When $Q=0.5$, $\langle V\rangle$ has a maximum($\langle V\rangle_{max}=1.65$) at $q=0.55$, and $\langle V\rangle\rightarrow0$ if $q\geq0.95$. When $Q=1$, $\langle V\rangle$ has a maximum($\langle V\rangle_{max}=1.70$) at $q=0.25$, and $\langle V\rangle\rightarrow0$ if $q\geq0.85$. So when $Q=0.5$(or $Q=1$), the particles get the fastest speed at $q=0.55$(or $q=0.25$), but this directional movement will disappear if $q\geq0.95$(or $q\geq0.85$). When $Q=1.5$, $\langle V\rangle$ decreases monotonically with increasing $q$, and the directional moment disappear($\langle V\rangle\rightarrow0$) if $q\geq 0.75$. As known, $\xi(t)$ coincides with the Gaussian colored noise in the limit of $q\rightarrow 1$, so in these cases, for the same value of noise intensity and the same value of correlation time, the non-Gaussian noise can induce particles directional movement, but Gaussian colored noise can not.

\begin{figure}
\center{
\includegraphics[height=8cm,width=10cm]{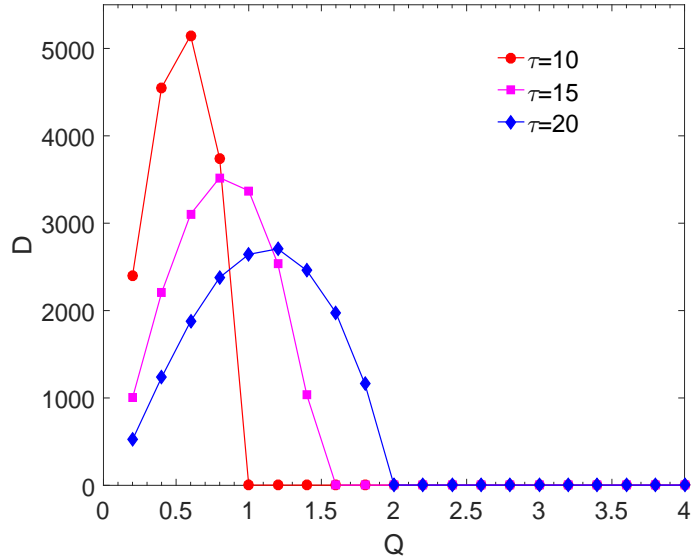}
\caption{The diffusion coefficient $D$  as a function of of noise intensity $Q$ for different values of $\tau$. The other parameters are $m=6$, $a=1.899$, $\omega=0.403$, $\varphi=\pi/2$, $q=0.8$. }\label{Fig4}
}
\end{figure}
Fig.(\ref{Fig4}) shows the diffusion coefficient $D$ as a function of $Q$ for different values of $\tau$. It is found that $D$ is a peaked function of $Q$(When $\tau=10$, $D_{max}=5148$ at $Q=0.6$. When $\tau=15$, $D_{max}=3520$ at $Q=0.8$. When $\tau=20$, $D_{max}=2706$ at $Q=1.2$). So for different values of noise intensity $Q$, there exits an optimal value of $\tau$ at which the particles achieve the best diffusion. In addition, the position of the peak shifts to large $Q$ when $\tau$ increases. In Fig.(\ref{Fig4}), we also find the particles diffusion should disappear($D\rightarrow 0$) if the value of $Q$ is large.

\begin{figure}
\center{
\includegraphics[height=8cm,width=10cm]{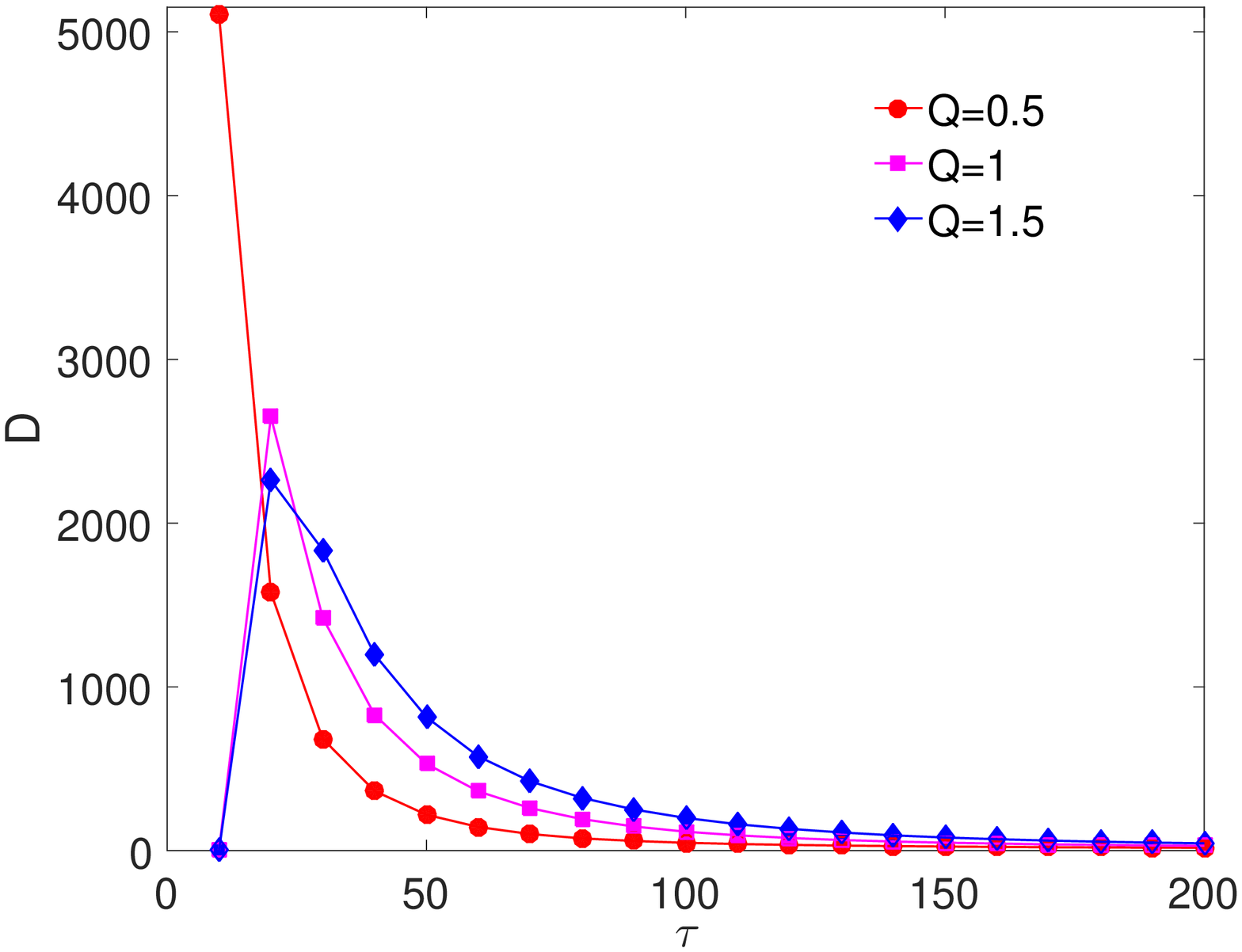}
\caption{The diffusion coefficient $D$  as a function of the self-correlation time $\tau$ for different values of noise intensity $Q$. The other parameters are $m=6$, $a=1.899$, $\omega=0.403$, $\varphi=\pi/2$, $q=0.8$.}\label{Fig5}
}
\end{figure}
Fig.(\ref{Fig5}) describes the diffusion coefficient $D$  as a function of $\tau$ for different values of $Q$. We find $D$ decreases  monotonically with increasing $\tau$ when $Q=0.5$. $D$ has a maximum with increasing $\tau$ when $Q=1.0$(or $Q=1.5$). So, when noise intensity is small($Q=0.5$), large $\tau$ will decreases the particles diffusion. But when noise intensity is large($Q=1.0$ and $Q=1.5$), there exists an optimal value of $\tau$ at which diffusion coefficient is maximal. Compare Fig.(\ref{Fig5}) and Fig.(\ref{Fig2}), we find $\tau$ has the same effect on $\langle V\rangle$ and $D$.

\begin{figure}
\center{
\includegraphics[height=8cm,width=10cm]{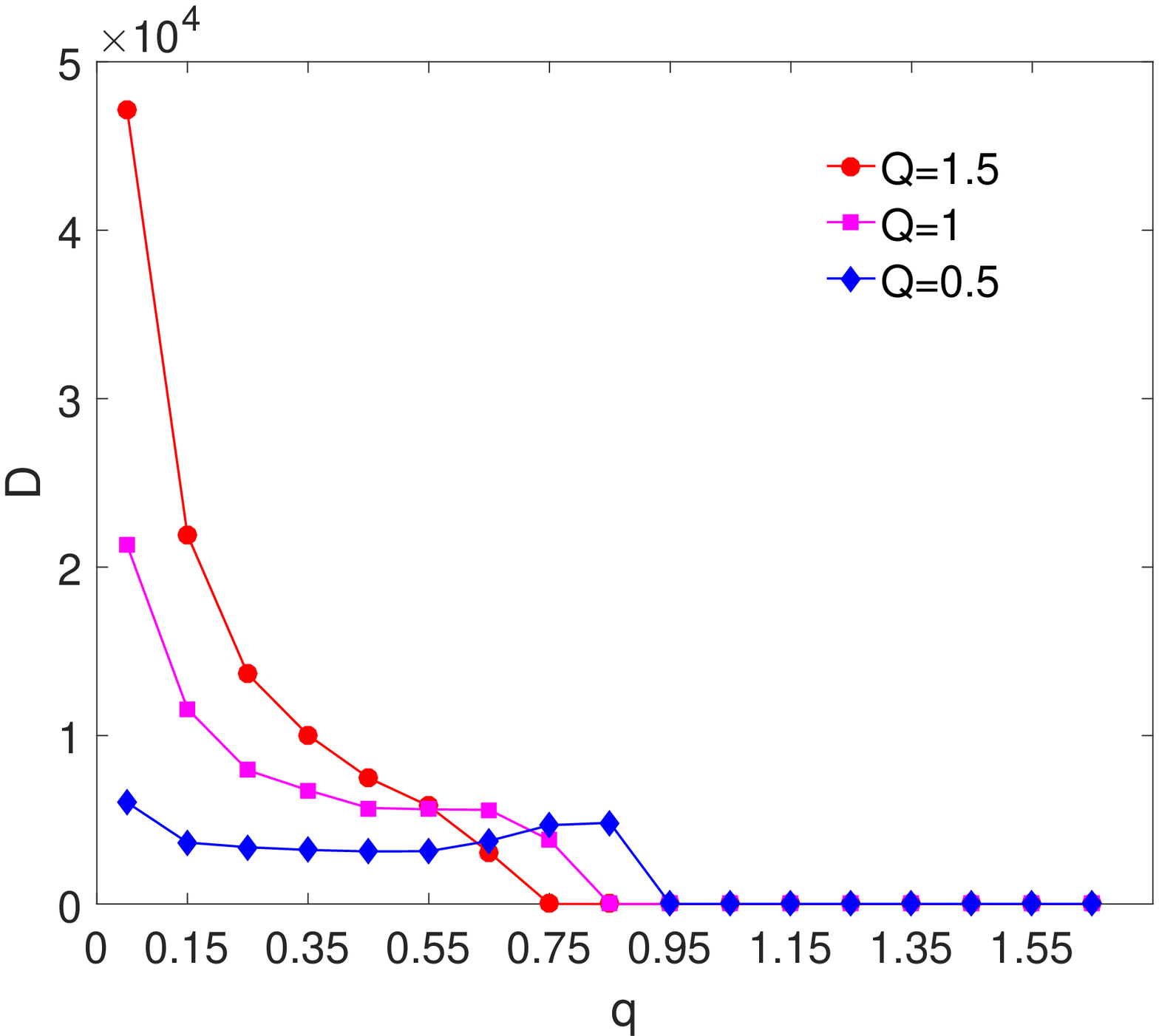}
\caption{The diffusion coefficient $D$ as a function of $q$ for different values of noise intensity $Q$. The other parameters are $m=6$, $a=1.899$, $\omega=0.403$, $\varphi=\pi/2$, $\tau=10$.}\label{Fig6}
}
\end{figure}
Fig.(\ref{Fig6}) shows $D$ as a function of parameter $q$ for different values of $Q$. When $Q=1$(or $Q=1.5$), $D$ decreases monotonically with increasing $q$, and $D\rightarrow0$ (the diffusion phenomenon disappear) if $q\geq0.85$(or $q\geq 0.75$). When $Q=0.5$, $D$ has a maximum($D_{max}=4821$) at $q=0.85$, and $D\rightarrow0$ if $q\geq0.95$. So, when $Q=0.5$, the particles get the best diffusion at $q=0.85$, but this phenomenon will disappear if $q\geq0.95$. Just like the results from Fig.(\ref{Fig3}), for the same value of $Q$ and the same value of $\tau$, non-Gaussian noise can induce particles diffusion, but Gaussian colored noise can not.

\section{\label{label4}Conclusions}
In this paper, we numerically studied the directional movement and diffusion of  particles in a periodic potential with non-Gaussian noise. We find that the average velocity $\langle V\rangle$(or the diffusion coefficient $D$) has a maximum with increasing noise intensity $Q$. When $Q=0.5$, the average velocity $\langle V\rangle$(or the diffusion coefficient $D$) decreases monotonically with increasing $\tau$. When $Q=1$(or $Q=1.5$), $\langle V\rangle$(or $D$) has a maximum with increasing $\tau$. In some cases, for the same value of $Q$ and the same value of $\tau$, the non-Gaussian noise can induce particles directional movement and diffusion, but the Gaussian colored noise can not.

\section*{Acknowledgments}
Project supported by Natural Science Foundation of Anhui Province(Grant No:1408085QA11).

\end{document}